%% file: article.tex
\documentclass{llncs}
\usepackage{german}
\usepackage{graphicx}
\usepackage{amsmath}
\usepackage{amssymb}
\usepackage{epic}
\usepackage{epsfig}
\usepackage{hyperref}
\usepackage[latin1]{inputenc}

\begin{document}
\mainmatter

\title{Koordinatenfreies Lokationsbewusstsein (Localization without Coordinates)}
\titlerunning{Koordinatenfreies Lokationsbewusstsein}
\author{ A.~Kr{\"o}ller\inst{1}\thanks{Supported by
DFG Focus Program 1126, ``Algorithmic Aspects of Large and Complex Networks'',
Grants Fe 407/9-1 and Fi 605/8-1.}
\and S.~P.~Fekete\inst{1} 
\and C.~Buschmann\inst{2}
\and S.~Fischer\inst{2} 
\and D.~Pfisterer\inst{2}\footnotemark[1]
}

\authorrunning{S.~P.~Fekete et al.}
\institute{%
 Institut f"ur Mathematische Optimierung\\
 Technische Universit"at Braunschweig\\
 D-38106 Braunschweig, Germany\\
 \email{\{s.fekete,a.kroeller\}@tu-bs.de}.\\
\and
 Institut f"ur Telematik\\
 Universit"at zu L"ubeck\\
 D-23538 L"ubeck, Germany\\
 \email{\{buschmann,fischer,pfisterer\}@itm.uni-luebeck.de}.
}

\authorrunning{S.~P.~Fekete et al.}
\tocauthor{S\'andor P.~Fekete (Braunschweig),
Alexander Kr\"oller (Braunschweig),
Dennis Pfisterer (L"ubeck),
Stefan Fischer (L"ubeck),
Carsten Buschmann (L"ubeck)}

\maketitle              

\begin{abstract}
Localization is one of the fundamental issues in sensor
  networks.  It is almost always assumed that it must be solved by
  assigning coordinates to the nodes. This article discusses
  positioning algorithms from a theoretical, practical and simulative
  point of view, and identifies difficulties and limitations.  Ideas
  for more abstract means of location awareness are presented and the
  resulting possible improvements for applications are shown.  Nodes
  with certain topological or environmental properties are clustered,
  and the neighborhood structure of the clusters is modeled as a
  graph.

\bigskip
  Eines der fundamentalen Probleme in Sensornetzwerken
  besteht darin, ein Bewusstsein f"ur die Position eines Knotens im
  Netz zu entwickeln.  Dabei wird fast immer davon ausgegangen, dass
  dies durch die Zuweisung von Koordinaten zu erfolgen hat. In diesem
  Artikel wird auf theoretischer, praktischer und simulativer Ebene
  ein kritischer Blick auf entsprechende Verfahren geworfen, und es
  werden Grenzen aufgezeigt. Es wird ein Ansatz vorgestellt, mit dem
  in der Zukunft eine abstrakte Form von Lokationsbewusstsein
  etabliert werden kann, und es wird gezeigt, wie Anwendungen dadurch
  verbessert werden k"onnen. Er basiert auf einer
  graphenbasierten Modellierung des Netzes: Knoten mit bestimmten
  topologischen oder Umwelteigenschaften werden zu Clustern
  zusammengefasst, und Clusternachbarschaften dann als Graphen
  modelliert.
\end{abstract}


{\bf Schlagworte:} Sensornetze, Lokalisierung, Cluster, Graphen, Routing



\input{commands}

\input{introduction}

\input{coordalgos}

\input{vision}

\input{summary}

\end{document}

%% file: commands.tex
\newcommand{\TOP}[1]{\paragraph{--- TOP: \glqq #1\grqq ---}\mbox{}\\}

\newcounter{TODOcnt}
\newcommand{\TODO}[1]{[{\em #1}]\stepcounter{TODOcnt}}

\newcommand{\PeqNP}{$\mathcal{P}=\mathcal{NP}$}
\newcommand{\NP}{$\mathcal{NP}$}
\newcommand{\NPv}{\NP-vollst"andig}
\newcommand{\NPvige}{\NP-vollst"andige}

\newcommand{\critnear}{C1}
\newcommand{\critfar}{C2}


%% file: introduction.tex
\section{Einleitung}
\label{sec:intro}

Sensornetze stellen einen vergleichsweise jungen Forschungsbereich
dar, in dem noch viele grundlegende Fragen ungekl"art sind. Eine
dieser Fragen ist, wie Sensorknoten ein Bewusstsein f"ur ihre Lage im
Netz entwickeln k"onnen, und wie dieses in Anwendungen ausgenutzt
werden kann. Dieses Bewusstsein kann beispielsweise darin bestehen,
dass jeder Knoten seine Koordinaten in einem gemeinsamen System kennt.
Da diese Form intuitiv sinnvoll erscheint, existiert eine F"ulle von
heuristischen Verfahren, die Koordinaten berechnen. Andererseits gibt
es derzeit praktisch keine alternativen Ans"atze. Ziel dieses Artikels
ist es, diese intuitive Entscheidung in Frage zu stellen und
Alternativen aufzuzeigen.

\subsection{"Uberblick}

Inhaltlich l"asst sich dieser Artikel durch drei Kernthesen
charakterisieren, die im Folgenden n"aher ausgef"uhrt werden.

\begin{itemize}
\item Die Zuweisung von Koordinaten stellt ein im
  komplexit"atstheoretischen Sinne schweres Problem dar, so dass es
  voraussichtlich keine fehlerfreien Algorithmen f"ur Sensornetze
  geben kann. (Abschnitt~\ref{sec:analyse:komplex})
\item G"angige Lokationsalgorithmen schlagen auch in realistischen
  Anwendungsszenarien fehl. (Abschnitt~\ref{sec:analyse:praxis})
\item Es ist m"oglich, Lokationsinformation zu generieren, die
  unabh"angig von Koordinaten ist. F"ur bestimmte Anwendungen ist
  dieses abstrakte Lokationsbewu"stsein dem klassichen Ansatz sogar
  "uberlegen. (Abschnitt~\ref{sec:vision})
\end{itemize}

Im Folgenden wird zun"achst in \ref{sec:scenario} ein
Anwendungsszenario vorgestellt, das durchgehend als Beispiel dient.
Abschnitt~\ref{sec:coordalgos} geht auf koordinatenbasierte
Lokationsalgorithmen ein. Dazu werden Klassifikationen f"ur Verfahren
vorgestellt und die Anforderungen an die zu berechnenden L"osungen
diskutiert. In Abschnitt~\ref{sec:analyse} wird das zugrunde liegende
Problem zun"achst theoretisch analysiert und auf Auswirkungen in der
Praxis eingegangen.  Au"serdem werden vier Algorithmen vorgestellt und
simulativ untersucht, um nachzuweisen, dass die beschriebenen Probleme
in der Realit"at tats"achlich auftreten. Abschnitt~\ref{sec:vision}
skizziert, wie neue Ans"atze genutzt werden k"onnen, um Alternativen
zu den bekannten Algorithmen zu entwickeln. Es wird ein Verfahren
umrissen, das topologisiche oder Umwelteigenschaften zum Clustern von
Knoten nutzt, und sich daraus ergebende Clusternachbarschaften mit
Hilfe eines Graphen modelliert.

\subsection{Szenario}
\label{sec:scenario}

In der Literatur ist es "ublich, Algorithmen auf einem Sensornetz zu
testen, bei dem die Knoten gleichverteilt auf einem konvexen Gebiet,
etwa einem Kreis oder Quadrat, platziert sind. Viele Verfahren
verwenden Anker, das hei"st Knoten, die ihre Koordinaten bereits im
Vorfeld kennen. Diese folgen im Allgemeinen derselben Verteilung wie
die "ubrigen \glqq normalen\grqq\ Knoten.  Viele anwendungsorientierte
Szenarien sind weniger gutartig. Wie wir im Folgenden zeigen, liefern
existierende Verfahren selbst unter leichten Abschw"achungen der
obigen Annahmen, insbesondere "uber die topologische Struktur des
betrachteten Gebiets, unbrauchbare Ergebnisse.

Im Folgenden wird ein derartiges Szenario verwendet. Darin sind die
Sensorknoten zuf"allig "uber ein Gebiet verteilt, das aus Stra"sen
besteht.  Es wird keine Gleichverteilung verwendet -- einige Stra"sen
sind dichter besetzt als andere. Ankerknoten sind nur in einigen
Bereichen vertreten. Wir sind davon "uberzeugt, dass dieser Aufbau
praktisch relevant ist, da es eine Vielzahl von Szenarien abdeckt.
Abgesehen von Stra"sen einer Stadt sind dies z.B.:
\begin{itemize}
\item Kanalisationen.
\item Von wenigen Fahrzeugen ausgebrachte Knoten.
\item "Uber einer Seenlandschaft abgeworfene Knoten.
\item Sensornetze, in denen in mehreren Gebieten alle Knoten
  ausgefallen sind und L"ocher erzeugt haben.
\end{itemize}

Abbildung~\ref{fig:input} zeigt das Sensornetz, das wir f"ur Tests
verwenden. Es sind nicht alle Kanten des Kommunikationsgraphen
dargestellt, sondern aus Gr"unden der "Ubersichtlichkeit nur ein
darunter liegender Teilgraph. Das Netzwerk besteht aus 2200 Knoten mit
durchschnittlich jeweils 50 Nachbarn. Die schwarzen Kreise markieren
die 200 Ankerknoten.

Die Ankerknoten befinden sich an zwei R"andern. Da
Lokationsalgorithmen oft die Position eines Knotens anhand einiger
umliegender Anker bestimmen, bilden sich zwischen Ankern zwangsl"aufig
Zellen mit Knoten, die anhand der selben Ankermenge verortet werden.
Da diese Zellen algorithmisch vollst"andig unabh"angig sind, ist es
sinnvoll, die Vorg"ange innerhalb einer solchen Zelle zu untersuchen.
Das verwendete Beispielnetz steht f"ur eine einzelne Zelle in einem
gr"o"seren Netzwerk, in dem Ankerknoten zuf"allig verteilt sind.

Dieser Artikel konzentriert sich auf Verfahren, die ein Sensornetz
selbst aus\-f"uhren kann. Algorithmen, die eine zentrale Recheninstanz
ben"otigen, werden daher nicht betrachtet. Fragen der Laufzeit und
Nachrichtenkodierung werden ebenfalls ignoriert, da es hier
ausschlie"slich um die G"ute der produzierten L"osungen geht.



%% file: coordalgos.tex
\section{Lokations\-be\-wusst\-sein durch Koordinaten}
\label{sec:coordalgos}


Das klassische Lokationsproblem besteht darin, jedem Knoten eine
Position im zwei- oder dreidimensionalen Raum zuzuordnen. Wenn ein
externes \glqq echtes\grqq\  Koordinatensystem vorgegeben ist, sollen
die errechneten Positionen den wirklichen m"oglichst genau
entsprechen. Im anderen Fall spricht man von virtuellen Koordinaten,
die nur Konsistenzbedingungen erf"ullen sollen.

Die Rahmenbedingungen f"ur ein Lokationsverfahren entsprechen denen,
die bei Sensornetzen "ublicherweise f"ur jeden Algorithmus gelten: Ein
Verfahren sollte
\begin{itemize}
\item m"oglichst wenig externe Infrastruktur ben"otigen,
\item mit m"oglichst wenig Kommunikation auskommen,
\item m"oglichst geringe Anforderungen an die Ressourcen der Knoten,
  etwa Prozessor und Speicher, stellen und
\item ohne externe Konfiguration im Rahmen des Ausbringens der Knoten
  auskommen.
\end{itemize}

Ein triviales Beispiel f"ur ein Positionierungsverfahren besteht also
darin, jeden Sensorknoten mit einem GPS-Empf"anger auszustatten. Der
Kommunikationsaufwand ist dabei optimal, da die Knoten keinerlei
Nachrichten austauschen.



\subsection{Klassifikation}

Lokationsverfahren lassen sich anhand mehrerer Kriterien
kategorisieren. Eine Gliederung wird nach dem Kriterium möglich, ob
ein Verfahren die Messung der Abstände zwischen benachbarten Knoten
vorsieht. So könnten die Knoten mit spezieller Hardware ausgestattet
sein, die die Abstandsbestimmung zu anderen Geräten in
Kommunikationsreichweite ermöglicht, z.B. mittels Ultraschall oder
Infrarot. Solche Abstandsmessungen sind jedoch im Allgemeinen mit
Messfehlern von bis zu 30\% behaftet
\cite{whitehouse,buschmannfischergiitg}. Während die eine Gruppe der
Verfahren solche Messungen vorsieht, um dennoch differenziertere
Abstandsinformationen in der Größenordnung unterhalb des
Kommunikationsradiuses einzusetzen, verzichtet die andere darauf und
beschränkt sich auf Abstandsabschätzungen mit Hilfe des
Kommunikationsradiuses.

Durch die Zahl der Ankerknoten wird eine weitere Unterteilung möglich.
Anker sind spezielle Knoten, die ihre Position kennen, z.B. durch GPS.
Sie dienen anderen Knoten als Referenzpunkte. Drei Gruppen von
Verfahren lassen sich so unterscheiden:
\begin{itemize}
\item Diejenigen, die völlig ohne Anker arbeiten, und die also nur
  eine konsistente Einordnung der Knoten in eine gemeinsames,
  virtuelles Koordinatensystem anstreben, 
\item sowie diejenigen, in denen jeder Knoten mehrere Anker mit
  bekannten Koordinaten in seinem Kommunikationsradius hat.
\item Dazwischen liegen die Verfahren, in denen die meisten Knoten
  keinen Anker direkt hören, sondern Nachrichten von Ankern über
  mehrere Hops zu den Knoten gelangen.
\end{itemize}

Die meisten auf Ankern basierenden Lokationsverfahren lassen sich in
drei Phasen unterteilen:
\begin{enumerate}
\item Bestimmung des Abstandes zu verschiedenen Ankern, d.h. zu
  bestimmten Koordinatenpunkten.
\item Berechnung der eigenen Position aus diesen Entfernungen und
  Punkten oder Winkeln dazwischen
\item Verfeinerung der eigenen Positionsschätzung durch lokale
  Heuristiken.
\end{enumerate}
In einigen Algorithmen fallen einzelne Phasen weg, insbesondere die
dritte. Andere verwenden ausschließlich diese, da sie sehr gut
parallelisierbar ist. Im Allgemeinen ist es möglich, die einzelnen
Phasen verschiedener Algorithmen auszutauschen
\cite{langendoen03comparison}.

\subsection{Konsistenz und Faltung}

Eine entscheidendes Kriterium, das eine Koordinatenzuweisung erf"ullen
sollte, ist Konsistenz. Darunter fallen die folgenden zwei
Forderungen:
\begin{enumerate}
\item[(\critnear)] Benachbarte Knoten bekommen nah beieinander
  liegende Koordinaten.
\item[(\critfar)] Nicht benachbarte Knoten bekommen Koordinaten, die
  einen Mindestabstand einhalten.
\end{enumerate}

Von einer Faltung spricht man, wenn Knoten, die in Wirklichkeit weit
entfernt sind, "ahnliche Koordinaten bekommen. Dadurch werden in der
r"aumlichen Struktur des Netzes zwei Bereiche \glqq "ubereinander
gefaltet\grqq.

Das erste Kriterium l"asst sich mit verteilten Algorithmen ohne
Schwierigkeiten "uberpr"ufen, und es gibt sehr viele verteilte
Verbesserungsheuristiken, wie beispielsweise Spring Embedder
\cite{priyantha03anchorfreelocalization}.

Das zweite Kriterium stellt im Kontext verteilter Systeme die
eigentliche Herausforderung dar. Es ist den Knoten im Allgemeinen
nicht m"oglich, die gesamte Netzwerkstruktur mit allen Koordinaten
abzuspeichern oder auf energieeffiziente Weise zu "ubertragen. Da die
Knoten also nicht wissen, ob nicht benachbarte Knoten dieselben
Koordinaten speichern, gibt es keinen einfachen Test, mit dem
Faltungen entdeckt werden k"onnen.

Viele Algorithmen setzen auf Ankerknoten, um Faltungsfreiheit zu
erzeugen. Oft wird im eigentlichen Verfahren ausschlie"slich das erste
Kriterium ber"ucksichtigt.  Dadurch, dass die Ankerknoten ihre
Position nicht ver"andern d"urfen, wird das zweite Kriterium implizit
eingebracht. Derartige Annahmen sind aber sehr fragil. Sie verlieren
ihre G"ultigkeit, wenn Abstandssch"atzungen tendenziell immer etwas zu
gro"s oder zu klein sind, es ankerfreie Gebiete im Netz gibt oder die
Positionsbestimmung der Anker ungenau ist.

\section{Problemanalyse}
\label{sec:analyse}

Der folgende Abschnitt untersucht drei wesentliche Punkte, die im
Zusammenhang mit Lokationsalgorithmen beachtet werden m"ussen: Die
theoretische Komplexit"at, praktische Erw"agungen und die Qualit"at
der von g"angigen Algorithmen produzierten Koordinaten.

\subsection{Komplexit"at}
\label{sec:analyse:komplex}

Gerade im Zusammenhang mit Konsistenz gibt es eine Reihe von
Negativresultaten, die theoretisch bewiesen sind.

In \cite{breu98unit} wird das \problemname{Unit Disk Graph Recognition
  Problem} untersucht. Dabei ist der Kommunikationsgraph ohne
Abstandssch"atzungen gegeben, und es wird eine konsistente
Koordinatenzuweisung gesucht. Daf"ur m"ussen Nachbarn unterhalb einer
frei w"ahlbaren Maximaldistanz $M$ und Nichtnachbarn mit einer
gr"o"seren Distanz platziert werden. Es wird gezeigt, dass bereits
dieses grundlegende Problem \NPv\ ist.

Ein M"oglichkeit, \NPvige\ Probleme anzugehen, liegt in
Approximationen. So kann zugelassen werden, da"s nicht benachbarte
Knoten nur einen Mindestabstand von $d\cdot M<M$ einhalten m"ussen.
In \cite{kuhn-udgapproximation} wird gezeigt, da"s dieses Problem f"ur
$d\geq \sqrt{2/3}+\varepsilon$ \NPv\ ist, wobei $\varepsilon$ f"ur
steigende Knotenzahlen gegen 0 konvergiert. Damit kann ein
polynomielles Approximationsschema nur existieren, wenn \PeqNP\ gilt.

Ein weiteres Problem ist \problemname{Unit Disk Graph Reconstruction}.
Hierbei wird nicht nur der Kommunikationsgraph vorgegeben, sondern
zus"atzlich das Ergebnis einer fehlerfreien Abstandsmessung
benachbarter Knoten. Gesucht ist eine Koordinatenzuweisung, die diese
Abst"ande einh"alt und nicht benachbarten Knoten einen gewissen
Mindestabstand zuweist. Auch hier kann wieder gezeigt werden, dass
dieses Problem \NPv\ ist \cite{aspnes04computational} und somit
voraussichtlich von einem Sensornetzwerk nicht zu l"osen ist.

Zu letzterem Problem existiert auch ein positives Resultat: In
\cite{eren04rigiditylocalization} wird eine Klasse von Netzwerken
identifiziert, f"ur die eine Lokalisierung mit polynomiellem Aufwand
m"oglich ist. Dabei mu"s unter anderem eine fehlerfreie
Abstandsmessung vorliegen, bei der s"amtliche zul"assigen L"osungen
durch Drehung und Translation auseinander hervorgehen. Zus"atzlich
wird gezeigt, dass die meisten Zufallsgraphen diese Eigenschaft haben.
Leider ist dieses Resultat nicht auf praktische Szenarien anwendbar,
da bereits kleinste Fehler in der Abstandsmessung die ben"otigten
Voraussetzungen zerst"oren und dadurch das Lokalisierungsproblem
wieder \NPv\ wird.

Insgesamt ist also davon auszugehen, dass auch weiterhin nur
Heuristiken zur Verf"ugung stehen, die unter ung"unstigen Umst"anden
inkonsistente Resultate produzieren.

\subsection{Praxis}
\label{sec:analyse:praxis}

Nach der theoretischen Untersuchung werden nun praktische Aspekte
diskutiert. Zuerst wird auf die f"ur Anker ben"otigten Ger"ate
eingegangen. Danach wird beispielhaft gezeigt, wie Inkonsistenzen in
der Positionierung Anwendungen empfindlich beeintr"achtigen k"onnen.

\subsubsection{Implementierung von Ankern:}

Die Verwendung von Ankerknoten bringt einige praktische Probleme mit
sich. Sie ben"otigen eine externe Quelle zur Bestimmung ihrer
Position. Typischerweise werden daf"ur GPS-Empf"anger eingesetzt.  Die
ben"otigte Hardware ist derzeit noch sehr gro"s, teuer und
energiehungrig. Es ist nach allgemeiner Auffassung nicht davon
auszugehen, dass zuk"unftige Hardware diese Probleme ausr"aumen kann.
Au"serdem ben"otigen GPS-Empf"anger Verbindungen zu geostation"aren
Satelliten. Damit scheiden sie f"ur den Einsatz unter der Erde, im
Meer, in Geb"auden oder auf anderen Planeten v"ollig aus. Somit sind
mehrere relevante Einsatzgebiete nicht abgedeckt, was zeigt, dass
ankerbasierte Verfahren f"ur viele wichtige praktische Zwecke nicht
ausreichend sind.


\subsubsection{Anwendung:}

Eine Beispielanwendung, die Positionierungsinformationen verwendet,
ist {\em GeoRouting}. Dabei werden Datenpakete nicht zu einem
speziellen Knoten geschickt, sondern an eine bestimmte Position.
Üblicherweise wird implizit vorausgesetzt, dass die
Lokationsinformation grundsätzlich in Form von Koordinaten vorliegt.
Eine Schwierigkeit, mit der viele der koordinatenbasierten {\em
  GeoRouting}-Verfahren zu kämpfen haben, liegt in dem Umstand
begründet, dass der geometrisch kürzeste Weg nicht unbedingt der beste
ist, gerade wenn Fragen der Energieeffizienz, Latenz oder
Ausfallsicherheit eine Rolle spielen. Die zweite offensichtliche
Schwierigkeit liegt in Faltungen. Ein Datenpaket kann hier wie bei
topologischen Löchern in einem lokalen Optimum enden, also bei einem
Knoten, der keinen Nachbarn mehr hat, der näher am Ziel liegt als er
selbst, vom Ziel aber noch weit entfernt ist. In solchen Situationen
werden verschiedene Heuristiken eingesetzt, wie z.B. in \cite{Blasi}
beschrieben. Ein denkbarer Ausweg könnte auch die lokale Verwendung
globalen Wissens sein.

Ein weiterer wichtiger Verwendungszweck von Lokationsinformationen
ist, mit Hilfe der Position des Knotens zusammen mit anderen
Informationen auf seinen Kontext zu schließen, und so von ihm
gelieferte Daten mit der Situation in Relation zu setzen. Wenn ein
Sensornetz beispielsweise dazu eingesetzt wird, ein bestimmtes
Ph"anomen zu "uberwachen und bei einem signifikanten Anstieg eines
Sensorwertes einen Alarm abzusetzen, kann eine Faltung katastrophal
sein: So w"urden etwa bei einer Mittelwertsberechnung "uber ein
geographisches Gebiet auch Knoten einbezogen, die weit entfernt liegen
und nur durch die Faltung "ahnliche Koordinaten haben. Dadurch
ver"andert sich der Mittelwert kaum, wenn das Ph"anomen nur im
angefragten Gebiet auftritt, und kann so f"alschlicherweise unter der
vorgegebenen Alarmschranke bleiben.

Für viele Anwendungen ist es wichtiger, ob Informationen von einer
abstrakten, symbolischen Lokation wie \glqq Kreuzung X\grqq\ stammen,
als von der Position (173,25; -35,89). Auch sollte der
Lokationsbegriff in der Lage sein, die Gruppierung von Knoten sowie
die lokationsbasierte Adaption von Algorithmen und Anwendungen zu
unterstützen.

Eine Diskussion grundlegender Lokationsmodelle sowie der Vorschlag
f"ur ein umfassendes Lokationsmodell, das symbolische und geometrische
Lokationen kombiniert, findet sich in \cite{leonhardt98location}. Das
dort beschriebene Modell eignet sich jedoch nicht f"ur den Einsatz in
Sensornetzwerken, da es aufgrunde seine Allgemeinheit sehr
schwergewichtig ist. Wir stellen daher in Abschnitt
\ref{sec:vision:georouting} eine alternative, koordinatenfreie Form
von Lokationsbewusstsein f"ur Sensornetzwerke vor, bei der die
verschiedenen Kritikpunkte und Anforderungen aufgegriffen werden.

\subsection{Simulation}

\begin{figure}
  \begin{center}
  \includegraphics[width=0.6\textwidth]{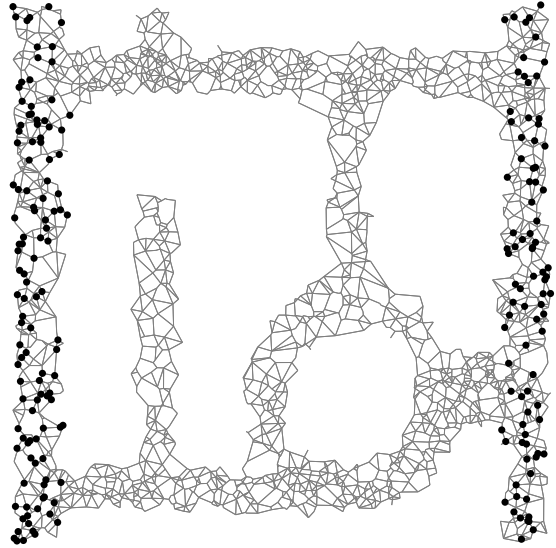}

  \caption{Beispielnetzwerk mit markierten Ankerknoten}
  \label{fig:input}
  \end{center}
\end{figure}

Um zu belegen, dass die beschrieben Probleme keine theoretischen
Sonderf"alle sind, die in \glqq realen\grqq\ Netzen nicht auftreten,
werden vier prominente Lokationsalgorithmen herangezogen. Wir
simulieren das Beispielszenario (Abbildung\ \ref{fig:input}) und
zeigen, dass tats"achlich Faltungen und Inkonsistenzen entstehen.

\subsubsection{Ein ankerbasierter zweiphasiger Algorithmus:}

{\em Ad-Hoc Positioning} \cite{niculescunath} ist ein Beispiel f"ur
einen zweiphasigen Algorithmus, der Ankerknoten und Distanzmessungen
verwendet. Mangels dritter Phase stehen einem Knoten die Koordinaten
seiner Nachbarn nicht zur Verf"ugung. Daher ist eine hohe Genauigkeit
der Sch"atzung von Entfernungen zu Ankerknoten entscheidend. Hierbei
wird, beginnend bei Ankerknoten, mittels Triangulierung "uber je zwei
oder drei Zwischenknoten die eigene Distanz zu Ankern bestimmt. Eine
perfekte Abstandsmessung und g"unstige Verteilung der Knoten
vorausgesetzt, ist diese Sch"atzung fehlerfrei.

Nachdem ein Knoten seine Entfernungen zu mehreren Ankern kennt,
berechnet er seine Position. Daf"ur wird {\em Multilateration}
verwendet, wobei ein quadratisches Gleichungssystem linearisiert und
dann eine Least-Squares-L"osung berechnet wird.  Bei perfekten
Messungen und g"unstigen Knotenverteilungen ergibt sich tats"achlich
die korrekte Position.

\begin{figure}
  \begin{center}
  \includegraphics[width=0.6\textwidth]{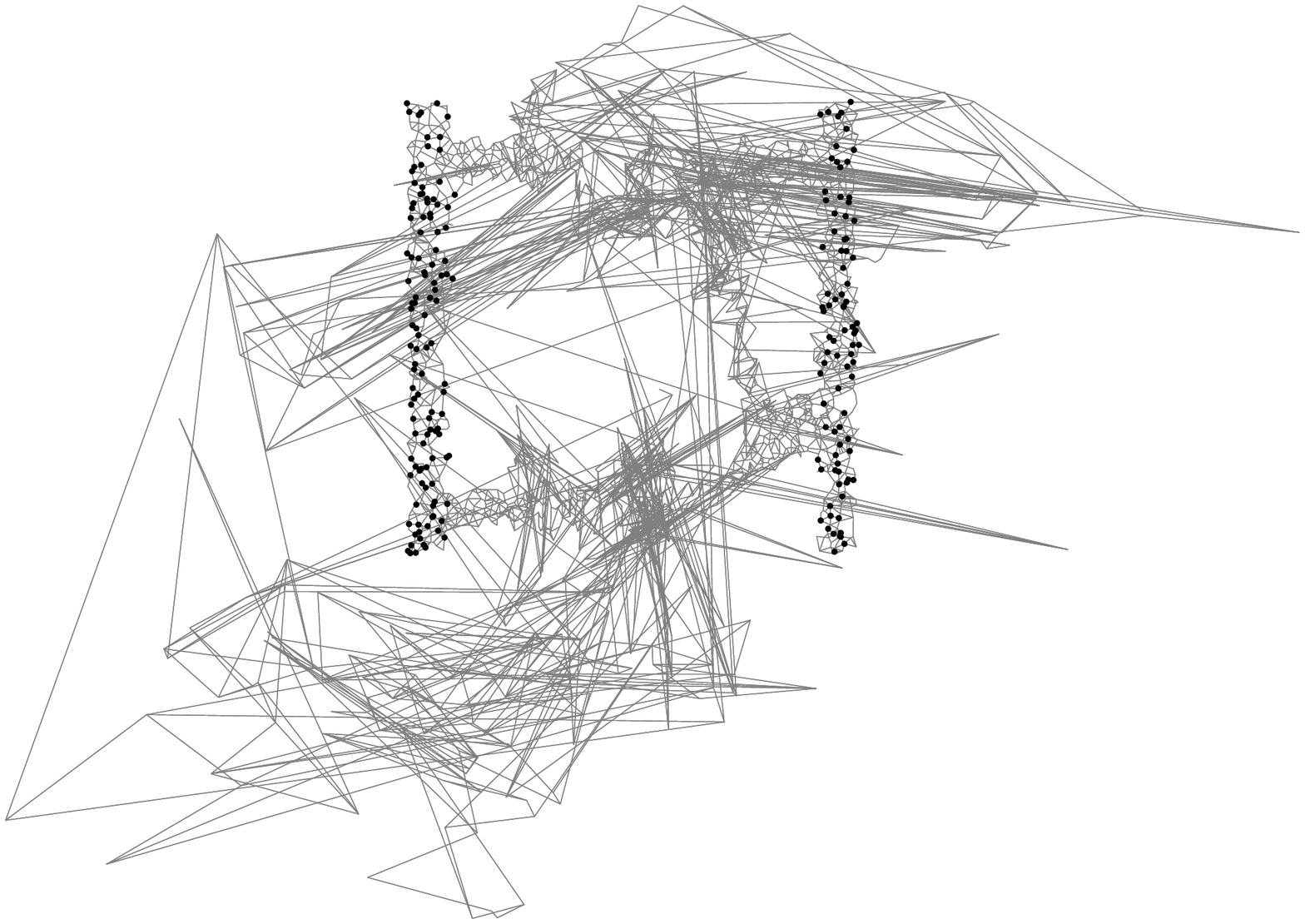}

  \caption{Ad-Hoc Positioning \cite{niculescunath}}
  \label{fig:euc-lat-none}
  \end{center}
\end{figure}

Abbildung\ \ref{fig:euc-lat-none} zeigt das Ergebnis dieses Verfahrens
bei einer Standardabweichung in Distanzmessungen von 1\%, also weit
weniger, als in der Praxis erreichbar ist. Es ist offensichtlich, dass
dieses Verfahren in der N"ahe der Anker gute Positionen berechnet;
allerdings summieren sich Fehler so schnell auf, dass schon nach
wenigen Schritten die berechneten Koordinaten v"ollig unbrauchbar
sind. Keines der beiden Kriterien \critnear\ und \critfar\ wird
erf"ullt.

\subsubsection{Ankerbasierte dreiphasige Algorithmen:}

Als Verbesserung bietet sich die Verwendung einer dritten Phase an.
Zwei prominente Vertreter dreiphasiger Algorithmen sind {\em Robust
  Positioning} \cite{savarese02robust} und {\em N-Hop Multilateration}
\cite{savvides02multilateration}. Bei beiden berechnen Knoten ihre
Position "uber wiederholte {\em Multilateration} zu ihren direkten
Nachbarn, um verteilt zu einer konsistenten L"osung zu konvergieren.

Die Unterschiede liegen im Wesentlichen in der Berechnung der
Startl"osung. {\em Robust Positioning} ben"otigt keine
Distanzmessungen, sondern verwendet {\em DV-Hop} in der ersten Phase.
Dabei bestimmen die Ankerknoten untereinander ein durchschnittliches
Verh"altnis von euklidischem Abstand und Hop-Abstand. Dieses Verh"altnis
verwenden die "ubrigen Knoten, um aus Hop- einen euklidischen Abstand
zu berechnen. Die Berechnung der Position l"auft danach "uber {\em
  Multilateration}. {\em N-Hop Multilateration} berechnet den Abstand
zu einem Anker als die Summe von Distanzmessungen "uber einen
k"urzesten Weg. Die Position wird berechnet, indem ein Knoten um jeden
bekannten Ankerpunkt ein Quadrat (bzw. einen W"urfel) mit der
Distanzsch"atzung als halber Kantenl"ange legt, und einen Punkt
bestimmt, der im Schnitt aller Quadrate liegt.

\begin{figure}
  \begin{center}
  \includegraphics[width=0.6\textwidth]{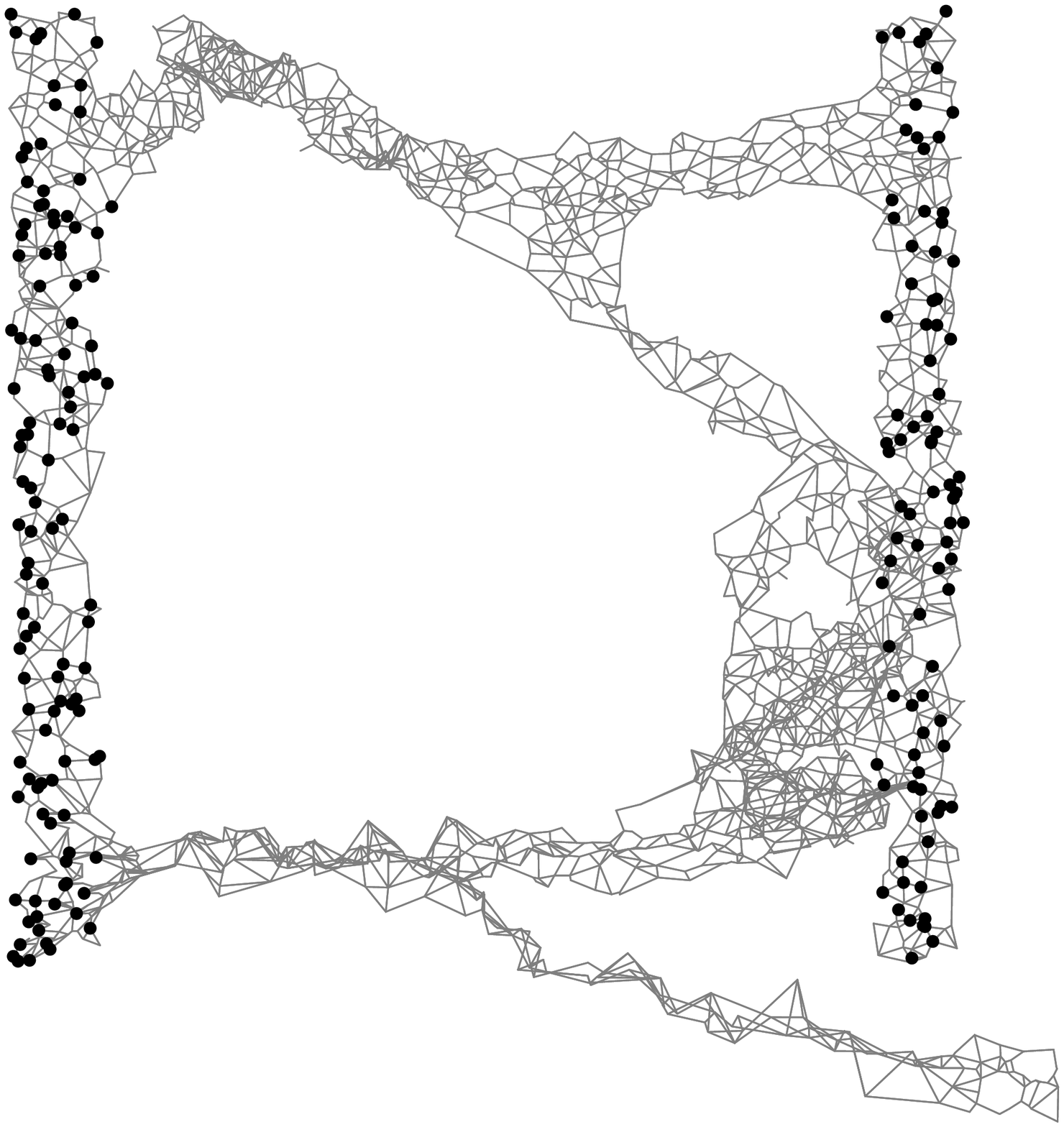}

  \caption{Robust Positioning \cite{savarese02robust}}
  \label{fig:dv-lat-ref}
  \end{center}
\end{figure}

\begin{figure}
  \begin{center}
  \includegraphics[width=0.6\textwidth]{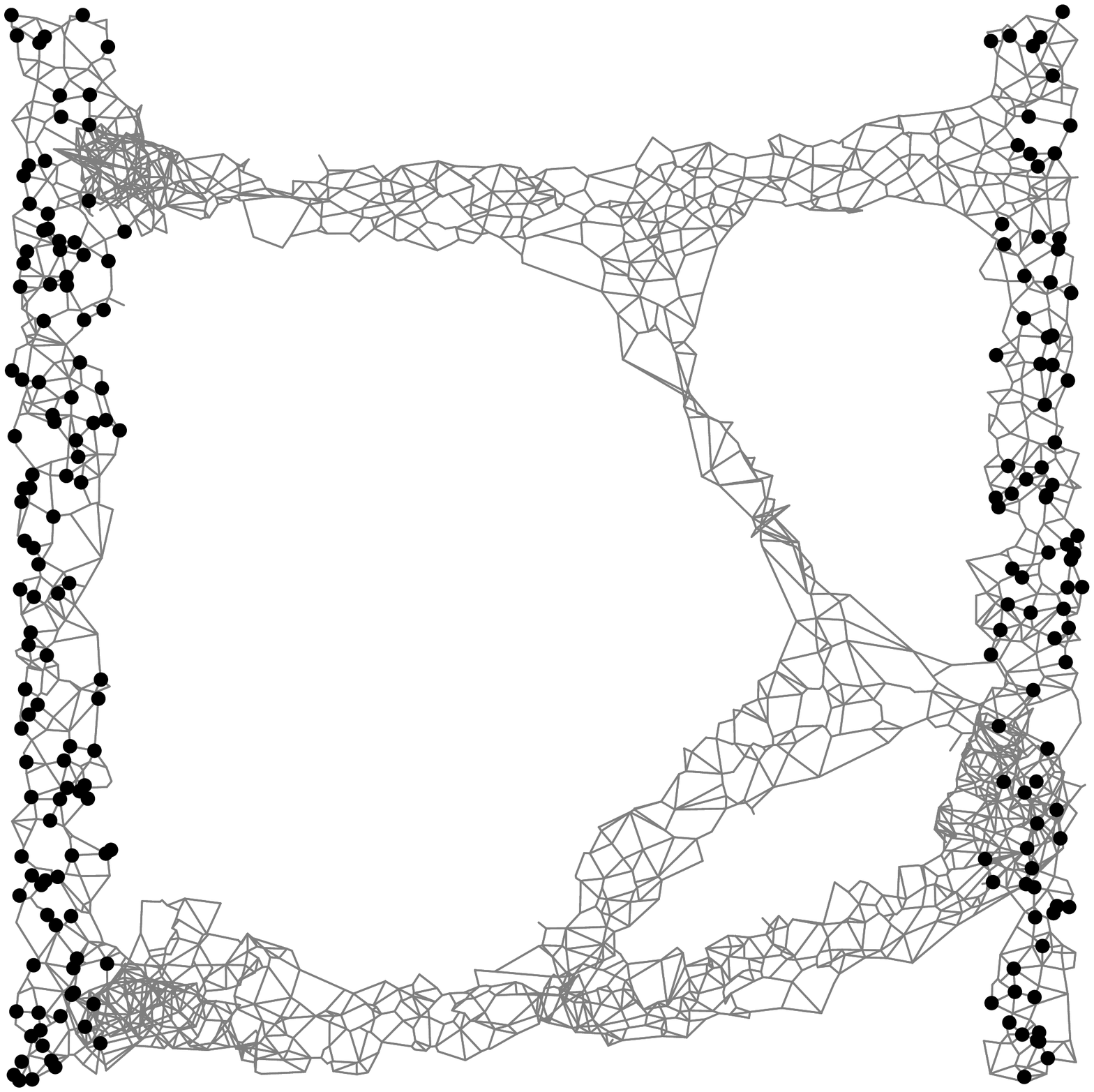}

  \caption{N-Hop Multilateration \cite{savvides02multilateration}}
  \label{fig:sum-box-ref}
  \end{center}
\end{figure}

Die berechneten L"osungen dieser beiden Algorithmen sind in den
Abbildungen~\ref{fig:dv-lat-ref} und \ref{fig:sum-box-ref}
dargestellt. Die dritte Phase eliminiert Ausrei"ser effektiv, wodurch
das Kriterium \critnear\ bei beiden Verfahren weitgehend erf"ullt
wird. Im Bereich der Ankerknoten ist der Positionierungfehler sehr
gering. Beide Verfahren haben massive Schwierigkeiten mit der
Sackgasse im Netzwerk. Da sie nicht berechnen k"onnen, in welcher
Richtung sie verl"auft, ist sie in beiden F"allen falsch plaziert.
Mangels direkter Ber"ucksichtigung des Kriteriums \critfar\ entstehen
hier fast zwangsl"aufig Faltungen. Diese Probleme verst"arken sich,
wenn die Knoten eine ungeschickte Auswahl an Ankern f"ur die
Positionierung treffen, oder die Verbindung zu Ankern nicht geradlinig
verl"auft, so dass Entfernungssch"atzungen sehr ungenau sind.

Diese Ergebnisse legen die Schlu"sfolgerung nahe, dass die Verwendung
von Ankern und entsprechenden Verfahren nur dann sinnvoll ist, wenn
sichergestellt werden kann, dass jeder Knoten hinreichend viele
Ankerknoten in n"aherer Umgebung hat. Sobald ankerfreie Bereiche
existieren, ist Faltungsfreiheit nicht mehr gew"ahrleistet.

\subsubsection{Ein ankerfreier Algorithmus:}

F"ur das Szenario ohne verf"ugbare Ankerknoten sind weniger
Algorithmen bekannt. {\em Anchor-Free Localization}
\cite{priyantha03anchorfreelocalization} sucht zun"achst heuristisch
zwei Knotenpaare. Dabei sind die Knoten jedes Paares m"og\-lichst weit
voneinander entfernt, und die Paare definieren zwei aufeinander
senkrecht stehende Achsen. Diese Achsen werden dann als
Koordinatenachsen verwendet, um eine m"oglichst faltungsfreie
Startl"osung zu erzeugen. Als Verbesserungsheuristik wird dann ein
verteilter Spring Embedder eingesetzt. Daf"ur stehen Distanzmessungen
zur Verf"ugung.

\begin{figure}
  \begin{center}
  \includegraphics[width=0.6\textwidth]{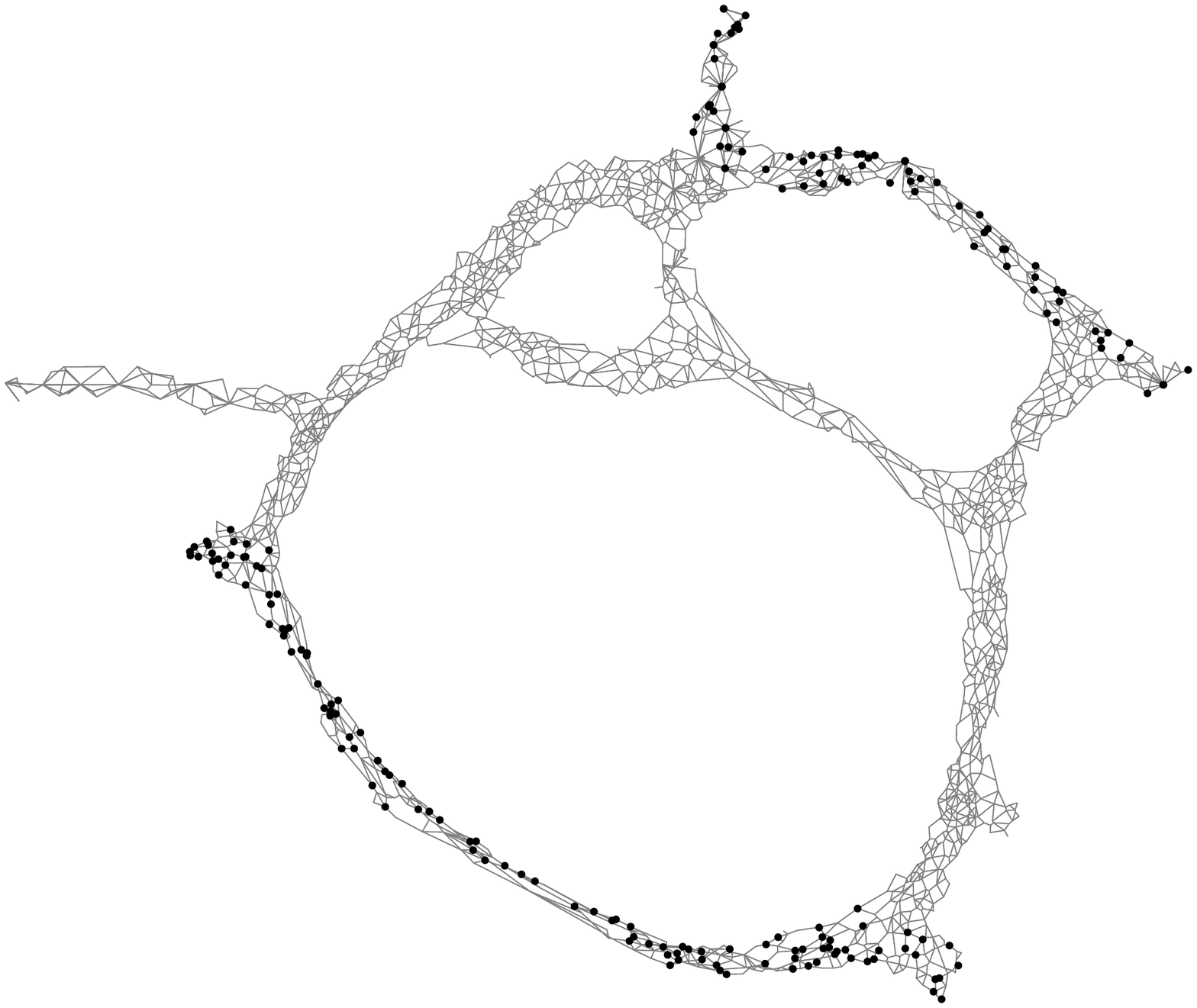}

  \caption{Anchor-Free Localization \cite{priyantha03anchorfreelocalization}}
  \label{fig:priyantha}
  \end{center}
\end{figure}

Abbildung~\ref{fig:priyantha} zeigt das Ergebnis dieses Verfahrens.
Da der Algorithmus ein eigenes Koordinatensystem verwendet, ist dieses
gegen"uber den vorigen L"osungen gedreht und gespiegelt. Abgesehen
davon ist die produzierte L"osung fast faltungsfrei und somit den
zuvor vorgestellten "uberlegen. Auch dieser Algorithmus produziert
Faltungen bei ung"unstigen Topologien, gerade in gr"o"seren Netzen;
das verwendete Beispielszenario l"ost er aber befriedigend. Es ist
also abzuwarten, welche Verbesserungen in zuk"unftige Versionen noch
erzielt werden k"onnen.

Da die Faltungsfreiheit nicht garantiert ist, sind die eingangs
angesprochenen Probleme auch hier nicht ausger"aumt. Im n"achsten
Abschnitt wird daher ein v"ollig neuer Ansatz vorgestellt.


%% file: vision.tex
\section{Abstraktes Lokationsbewusstsein}
\label{sec:vision}

Im Folgenden soll eine alternative Form von Lokationsbewusstsein
vorgestellt werden. Unseres Wissens nach ist sie trotz ihres gro"sen
Potenzials kaum erforscht. Die einzelnen Bestandteile sind dabei seit
langem bekannt und untersucht: Symbolische Lokation, topologische oder
geometrische Cluster zur Abstraktion der Topologie und gewichtete
Graphen zur Abbildung des Sensornetzes. Da es sich hierbei um
unerforschtes Neuland handelt, befinden sich entsprechende Verfahren
noch in der Entwicklungsphase.


Auf unterster Ebene werden Graphen zur Modellierung von Netzwerken
genutzt: Jeder Sensorknoten wird als Knoten modelliert. Kanten
verbinden je zwei Knoten, die miteinander direkt kommunizieren
k"onnen.

Die Idee besteht nun darin, Sensorknoten zu gro"sen Clustern zu
gruppieren und jeden Cluster als einzelnen Knoten des Graphen
aufzufassen. Clusternachbarschaften erzeugen dann die Kanten.  Der
Vorteil ist, dass die resultierenden Datenstrukturen sehr viel kleiner
als vollst"andige Kommunikationsgraphen sind.

Diese reduzierten Graphen k"onnen dann an alle Sensorknoten verteilt
werden. Das Wissen um den eigenen Cluster und seine Lage im Graphen
ist dann eine Art von Lokationsbewusstsein.

\subsection{Cluster und Topologie}

Ein wichtiger Schritt besteht darin, eine Clusterstruktur aufzubauen,
die die topologische Struktur des Netzes gut abbildet.

Im bereits erw"ahnten Beispiel des Stra"sensystems besteht ein gutes
Clustering etwa darin, jede Stra"se und jede Kreuzung zu jeweils einem
Cluster zusammenzufassen. Dazu muss das Netzwerk die lokale Topologie
erkennen k"onnen. Eine M"oglichkeit daf"ur wurde in
\cite{fekete04topbased} skizziert. Hier entscheiden die Knoten anhand
eines lokalen Kriteriums, ob sie am Rand des Netzes liegen, und bilden
zusammenh"angende Randstreifen. Eine Stra"se ergibt sich dann als
zusammenh"angende Knotenmenge, die demselben Randpaar folgt.
Kreuzungen sind dann Bereiche, in denen mehrere Stra"sen
zusammentreffen.

Besonders wichtig ist dabei, dass es keine Rolle spielt, ob die
Stra"senerkennung perfekt funktioniert. Entscheidend ist nur, dass die
Cluster die wesentliche topologische Struktur wiedergeben. Hierf"ur
existieren derzeit nur erste Ans"atze, die noch weiterentwickelt
werden m"ussen.

Eine zweite M"oglichkeit verwendet die Sensoren der Knoten. Dabei
schliessen sich Knoten zusammen, wenn sie in einer "ahnlichen Umgebung
liegen. Sofern die Knoten in der Lage sind, mit ihren Sensoren
abstrakte Umgebungseigenschaften wie beispielsweise \glqq Wasser\grqq,
\glqq Acker\grqq\ oder \glqq Schatten\grqq\ zu erkennen, kann eine
Clusterstruktur aufgebaut werden, die auf nat"urliche Weise die
Umgebung des Netzes abbildet.  Allerdings liegen auch hier noch sehr
viele ungel"oste Probleme und Fragestellungen, die erforscht werden
m"ussen.

\subsection{Cluster und Graphen}

Die Clusterstruktur wird in einen Graphen "uberf"uhrt, indem f"ur
jeden Cluster ein Knoten und f"ur benachbarte Cluster Kanten
eingef"uhrt werden. Alternativ kann, wenn die skizzierten Cluster des
Stra"senszenarios verwendet werden, ein Knoten f"ur jede Kreuzung und
eine Kante f"ur jede Stra"se erzeugt werden.

Ein hinreichend grobes Clustering vorausgesetzt, ist der entstehende
Graph klein genug, um ihn an alle Sensorknoten zu verteilen. Da die
Knoten wissen, in welchem Cluster sie sich befinden, bekommen sie
dadurch ein Bewusstsein "uber die Struktur des Netzes sowie ihre
eigene Position darin.

\begin{figure}
  \begin{center}
  \includegraphics[width=0.5\textwidth]{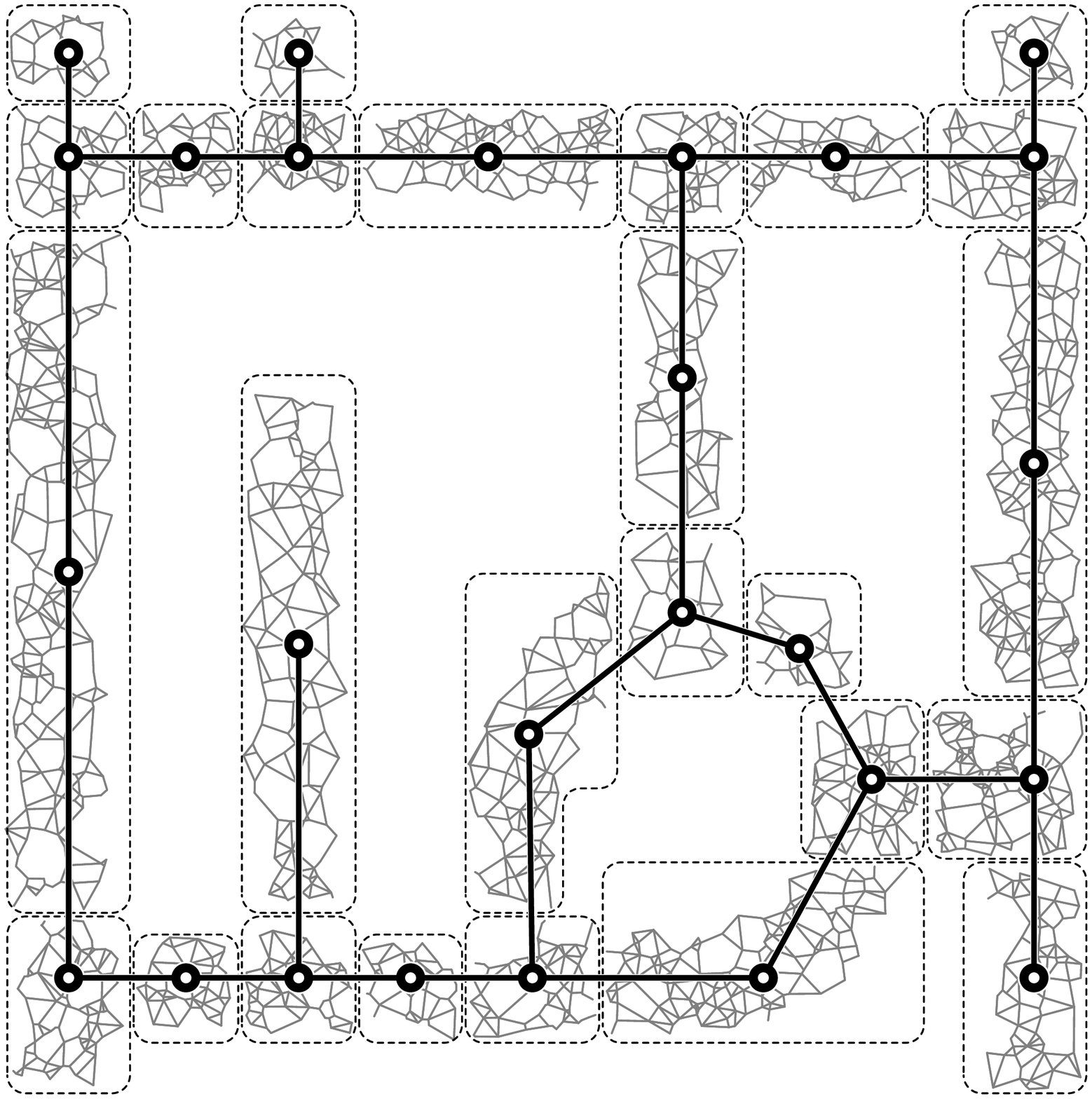}

  \caption{Cluster und Clustergraph}
  \label{fig:clustergraph}
  \end{center}
\end{figure}

Abbildung~\ref{fig:clustergraph} zeigt anhand des Stra"senbeispiels,
welche Sensorknoten sich zu Clustern zusammenschlie"sen, und welcher
Clustergraph daraus entsteht.

\subsection{Anwendung}
\label{sec:vision:georouting}

Da die Knoten "uber den erzeugten Graphen eine globale Sicht auf das
Sensornetz haben, k"onnen in jedem Bereich, in dem "ublicherweise
lokale Heuristiken zur Erreichung globaler Ziele eingesetzt werden,
Verbesserungen erzielt werden.  Beispielhaft soll dies im Folgenden
anhand von {\em GeoRouting} diskutiert werden.


Es gibt schon seit einiger Zeit Routingverfahren, die auf
hierarchische Cluster aufsetzen. Ein "Uberblick findet sich z.B. in
\cite{perkins-adhocnet}. Viele dieser Verfahren sind in Verruf
geraten, weil sie trotz des vorhandenen globalen Wissens nur auf
lokale Greedyheuristiken setzen. Dadurch kann es passieren, dass zwei
Knoten, die nahe beieinander liegen, aber in verschiedenen Clustern
sind, nur "uber sehr gro"se Umwege kommunizieren d"urfen. Dennoch sind
Cluster -- insbesondere die hier verwendeten topologischen -- sehr gut
geeignet, um mit lokalen Entscheidungen global g"unstige Routingpfade
aufzubauen.

Die Vorteile ergeben sich daraus, dass bereits der Sender die globale
Topologie zur Verf"ugung hat. Daher kann er den Routingpfad als
Sequenz benachbarter Cluster festlegen. Dabei k"onnen zus"atzliche
Informationen "uber den Graphen f"ur weitere Verbesserungen genutzt
werden.

So k"onnten die Kanten mit einer Sch"atzung der Restenergie oder
Kommunikationsbandbreite der Sensorknoten im zugeh"origen Cluster
abgespeichert werden. Dadurch kann bereits im Vorfeld ein
energieeffizienter Pfad hoher Bandbreite angelegt werden.  Wenn zu
erwarten ist, dass viele Daten zu "ubertragen sind, k"onnen
Load-Balancing-Strategien "uber disjunkte Pfade implementiert werden.
Dadurch wird gleichzeitig ausfallsichere Kommunikation erm"oglicht.

Auf Basis einer solchen Graphenmodellierung lassen sich viele weitere
Vorteile für die Anwendungen erzielen. So lassen sich beispielsweise
Clustern Eigenschaften zuordnen, die wiederum eine Kategorisierung
möglich machen. Diese kann dann als Kontextinformation genutzt werden.
So ließen sich beispielsweise aufgrund der Nachbarschaftsbeziehungen
Eigenschaften wie \glqq Kreuzung\grqq\ für einzelne Cluster ableiten.
Allgemein unterstützt die Clusterbildung die Kontextgewinnung im Netz,
beispielsweise über die relative globale Clusterposition am Rand oder
in der Mitte des Netzes. Auch ließen sich so Cluster abstrakteren
Ordnungsrelationen wie Reihenfolgen für Flussbetrachtungen im
Gesamtnetz unterwerfen. Auch Datenfusion lässt sich nicht nur auf
Basis von Koordinaten durchführen, sondern auch clusterbasiert, da
häufig Daten von Sensorknoten in ähnlichen Kontexten zusammengeführt
werden sollen.


%% file: summary.tex
\section{Zusammen\-fassung}

\label{sec:summary}

In diesem Artikel wird die in der aktuellen Forschung verfolgte
Vorgehensweise hinsichtlich der Lokalisierung diskutiert. 

Es wird aufgezeigt, dass oft ins Feld gef"uhrte Annahmen und
Voraussetzungen "uber die Gutartigkeit von Szenarien zu optimistisch
sind. Dies f"uhrt in anwendungsorientierten Szenarien dazu, dass es
vielmehr schwierig bis unm"oglich ist, zu konsistenten
Koordinatenzuweisungen zu gelangen.  Es wird theoretisch wie simulativ
belegt, dass ohne technische Hilfsmittel konsistente L"osungen
voraussichtlich nicht berechenbar sind. Nicht nur bei Anwendungen wie
beispielsweise dem GeoRouting k"onnen solche Inkonsistenzen jedoch zu
erheblichen Problemen führen, so dass die Suche nach alternativen
Ans"atzen angeraten ist.

Die Autoren schlagen daher vor, das Problem gerade in komplizierten,
realit"atsnahen Szenarien auf eine neue Art und Weise anzugehen. Wir
zeigen, wie Algorithmen mit Hilfe von geographischen Clustern und dem
zugeh"origen Verbindungsgraphen in der Lage sind, eine abstrakte Form von
Lokationsbewusstsein aufzubauen, die von externer Infrastruktur
vollst"andig un"abh"angig ist. Au"serdem ist dieser aggregierte
Graph so klein, dass er an alle Knoten im Netz verteilt werden kann.
Anders als beim koordinatenbasierten Ansatz erlangt somit jeder Knoten
ein globales Bild von der Netzstruktur.

Zur Umsetzung dieser Ans"atze sind noch viele Fragen offen, da der
Forschungsbereich kaum erschlossen ist. An der TU Braunschweig und der
Universit"at zu L"ubeck wird zur Zeit an verschiedenen Methoden zur
Errichtung abstrakten Lokationsbewusstseins gearbeitet. Es ist daher
zu erwarten, dass sich in naher Zukunft noch viele Erkenntnisse
ergeben und Verfahren entwickelt werden.
